\documentclass[letterpaper,twocolumn,10pt]{article}
\usepackage{usenix2019_v3}

\usepackage{tikz}
\usepackage{amsmath}
\usepackage{caption}
\usepackage{tabularx}

\usepackage{filecontents}

\begin{document}

\date{}

\title{WikipediaBot: Automated Adversarial Manipulation of Wikipedia Articles}

 \author{
 {\rm Filipo Sharevski}\\
 Divergent Design Lab \\ DePaul University 
 \and
 {\rm Peter Jachim}\\
 Divergent Design Lab \\ DePaul University
 } 

\maketitle

\begin{abstract}
This paper presents an automated adversarial mechanism called WikipediaBot. WikipediaBot allows an adversary to create and control a bot infrastructure for the purpose of adversarial edits of Wikipedia articles. The WikipediaBot is a self-contained mechanism with modules for generating credentials for Wikipedia editors, bypassing login protections, and a production of contextually-relevant adversarial edits for target Wikipedia articles that evade conventional detection. The contextually-relevant adversarial edits are generated using an adversarial Markov chain that incorporates a linguistic manipulation attack known as MIM or malware-induced misperceptions. Because the nefarious use of WikipediaBot could result in harmful damages to the integrity of wide range of Wikipedia articles, we provide an elaborate discussion about the implications, detection, and defenses Wikipedia could employ to address the threat of automated adversarial manipulations and acts of Wikipedia vandalism.
\end{abstract}

\section{Introduction}

The adversarial manipulation of publicly available information, such as Wikipedia, is an effort intended to deliberately degrade articles either by adding damaging or ill-intent content \cite{Tran}. In most cases, these edits consist of erasing the full text of articles, adding profanity or racial slurs, or editing the page form \cite{Mola-Velasco}. This nefarious form of manipulation is also known as \textit{Wikipedia vandalism} and often involves the use of unexpected words to draw attention \cite{Chin}. The vandals can be \textit{sneaky} and difficult to find because they make subtle edits and conceal their changes, aiming to deceive other editors and moderators into thinking the changes are legitimate \cite{Nguyen}. With these subtle changes, the adversary aims to break the consistency of articles, deviate from common or correct grammatical structure, introduce uncommon word patterns, or change the meaning of a sentence. 

A plethora of methods have been developed within the Wikipedia and the scientific community to detect Wikipedia vandalism including controversy detection on page level \cite{Aniket},\cite{Bykau}, analysis of deletions, insertions, number of revisions and text reliability \cite{Vuong}, \cite{Druck}, and analysis of author behaviour \cite{Rad}, \cite{Druck}. All of these methods assume that the adversary, in a role of a malicious editor (or group of editors), \textit{manually} vandalizes Wikipedia articles. To evade detection while still retaining the capability to make sneaky changes, we explored the possibility for an \textit{automated} adversarial manipulation of Wikipedia articles. We developed an automated adversarial mechanism called \textit{WikipediaBot} that uses a Markov chain and a linguistic manipulation paradigm called Malware Induced Misperceptions (MIM) to use word patterns from related articles to make malicious edits \cite{Sharevski1}. These malicious edits are made using a bot infrastructure and make contextual sense for the targeted content in the original article that doesn't decrease the text reliability or show an abnormal editor behaviour. 

We performed a proof-of-concept analysis of the WikipediaBot to understand the plausibility of the automated adversarial manipulation as an emerging threat to the credibility on online information. To this objective, the paradigm behind the WikipediaBot is described in Section~\ref{sec:2}. Section~\ref{sec:3} details the design and implementation of the WikipediaBot mechanism. The detection and the possible defenses against the WikipediaBot mechanism are discussed in Section~\ref{sec:4}. Section~\ref{sec:4} discusses the enhancements, limitations, and implications of using the WikipediaBot and Section~\ref{sec:6} concludes the paper. 

\section{WikipediaBot: Paradigm}
\label{sec:2}

\subsection{Malware-Induced Misperceptions}
The malware-induced misperception as a concept stems from the effort to explore alternative ways of instilling negative sentiment and divisiveness through linguistic manipulation of social media content \cite{Sharevski1}. The difference in the MIM method is that the manipulation is not made by directly crafting a malicious editor, but instead, by a man-in-the-middle malware that intercepts an original content, rearranges the words and text, and presents it to users to induce misperceptions. This misperception is a highly improbable interpretation of the set of true facts in the original content to the objective of the malicious actor \cite{Benkler}. So far, the MIM method has been tested in a browser extension variant as a method for conveying divisiveness in Facebook debates on freedom of speech \cite{Sharevski1} and Twitter vaccine debates \cite{Sharevski2}. The MIM browser extension manipulated keywords like "conservative zealots" with "far-right," "people against immunisation" with "anti-vaxxers," as well as replacing trolling hashtags like "\#chinavirus" with "\#covid19," among some of the examples. 

\subsection{Automated Adversarial Manipulation}
In the initial MIM work, the adversary manually inferred the linguistic manipulation of the content in the context of the social media discourse. This limited a wider scalability of the malware to produce contextually relevant word patterns as means of automated adaptations to evade human detection. Therefore, we extended the MIM paradigm to allow for automated \textit{context-relevant word substitutions}. We created an Markov chain algorithm for crafting context-relevant word substitutions in a candidate text for an adversarial manipulation. To describe our algorithm, we will use a minimal text example from the first verse of the poem “Jabberwocky” from \textit{Through the Looking Glass} by Lewis Carroll \cite{carroll}: 

\begin{table}[h]
\centering
    \begin{tabular}{c} 
   ‘Twas brillig, and the slithy toves \\
        Did gyre and gimble in the wabe; \\
        All mimsy were the borogoves, \\
        And the mome raths outgrabe.
    \end{tabular}
\end{table}

We divided the stanza into two pieces, a training set and a target adversarial manipulation set. Using the first two lines, we prepared a Markov chain with an initial and a final state, along with the probability that the initial state will be followed by the final state, as shown in Table 1. Next, we selected target words employing the MIM approach. In our case, we targeted occurrences of the words “borogoves” and “mome”:

\begin{table}[h]
\centering
    \begin{tabular}{c} 
        All mimsy were the TARGET, \\
        And the TARGET raths outgrabe. \\
    \end{tabular}
\end{table}

The word preceding the target word each time is the word “the.” When we look in Table 1, we see that there’s a 50\% chance that the word “the” is followed by the word “slithy”, and a 50\% chance that the word “the” is followed by the word “wabe.” We did a weighted random choice using the “probability” vector, transforming the verse to read like this:

\begin{table}[h]
\centering
    \begin{tabular}{c} 
        ‘Twas brillig, and the slithy toves \\
        Did gyre and gimble in the wabe; \\
        All mimsy were the wabe, \\
        And the slithy raths outgrabe. \\
    \end{tabular}
\end{table}

While this doesn’t fit the poetic structures in place, someone who is not familiar with “Jabberwocky” might not notice the changes, and the nonsensical vocabulary looks like it belongs. Running the Markov Chain MIM on the same passage of text using the same input text multiple times might yield different results, however, the poem still has a similar cadence:

\begin{table}[h]
\centering
    \begin{tabular}{c} 
        ‘Twas brillig, and the slithy toves \\
        Did gyre and gimble in the wabe; \\
        All mimsy were the slithy, \\
        And the slithy raths outgrabe. \\
    \end{tabular}
\end{table}

\begin{table}[h]
\caption{Markov Chain MIM: States and Probabilities} 
\centering
    \begin{tabular}{|c|c|c|} 
        \hline
         \textbf{Initial State} & \textbf{Final State} & \textbf{Probability}  \\
         \hline     
         ‘Twas & brillig, & 1 \\
         brillig, & and & 1 \\
         and & the & 0.5 \\ 
         the & slithy & 0.5 \\
         slithy & toves & 1 \\
         toves & Did & 1 \\
         Did & gyre & 1 \\
         gyre & and & 1 \\
         and & gimble & 0.5 \\
         gimble & in & 1 \\
         in & the & 1 \\
         the & wabe; & 0.5 \\
         \hline
    \end{tabular}
\end{table}

\section{WikipediaBot: Design and Implementation}
\label{sec:3}
\subsection{Design}
At a high level, the WikipediaBot mechanism consists of several modules as shown in Figure 1. The \textit{Generate Logins} module generates user credentials necessary for creating the adversarial manipulation bot structure. The login credentials are created by generating a random username and a random password as shown in Figure 2. To make the random username look more realistic, the \textit{Login Creator} module uses a random first name and a random last name from Wikipedia's lists of common first and last names, along with the password, which is simply a series of random characters \cite{surnames}, \cite{names}. The \textit{generate logins} module passes the credentials to the account creation page on Wikipedia. This page is scraped, and the bytes from the CAPTCHA are shown in the adversary's console, so the adversary only needs to enter the CAPTCHA details once (an example is provided in Figure 3). It takes about ten seconds to create each bot account.  

\begin{figure*}[h]
  \centering
  \includegraphics[width=0.6\linewidth]{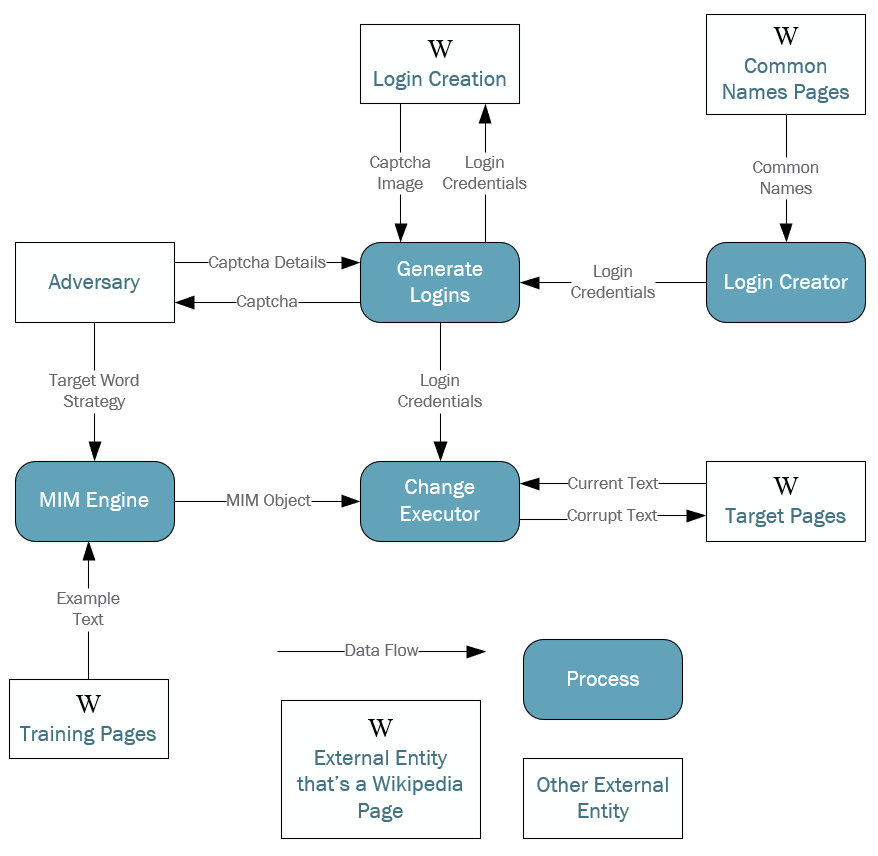}
  \caption{Data Flow Diagram of Application of MIM}
\end{figure*}

\begin{figure}[h]
  \centering
  \includegraphics[width=0.7\linewidth]{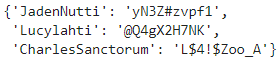}
  \caption{Sample Login Credentials}
\end{figure}

\begin{figure}[h]
  \centering
  \includegraphics[width=0.5\linewidth]{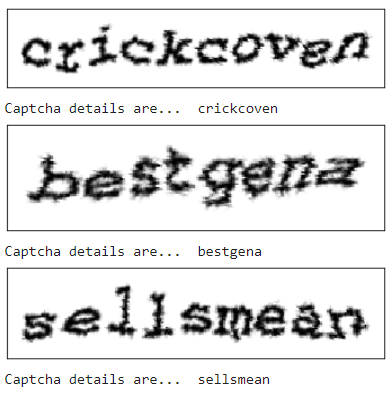}
  \caption{Captcha Inputs}
\end{figure}

The \textit{MIM Engine} module implements the Markov Chain MIM logic described in Section~\ref{sec:2}. As inputs, the \textit{MIM Engine} receives the target word strategy from the adversary and collects example texts from target Wikipedia articles. To make adversarial edits in the text, the \textit{MIM Engine} reads each word in a text, and based on the adversarial target word MIM strategy, replaces that word using a context-relevant word using the preceding two words in a Markov chain trained on example text from a training set of Wikipedia articles. The adversarial manipulation edits are then passed as a "MIM Object" to the \textit{Change Executor} module. The \textit{Change Executor} takes the MIM object, which caries all the context-relevant edits, logs as an editor to the target Wikipedia page and replaces the original content with the manipulated one. Except the \textit{MIM Engine}, each of the modules of WikipediaBot uses the python API of the Selenium library, which allows WebUI interaction indistinguishable from a regular human editor interaction with the Wikipedia pages \cite{selenium}. 

\subsection{Implementation Example}
To show how the WikipediaBot could be used to harm discourse, we analyzed a scenario where a hypothetical adversary aims to reduce mentions of Uyghurs on the Uyghurs Wikipedia page, shown in Table 2 \cite{wiki_uyghurs}. Uyghurs are a Muslim Chinese ethnic minority, many of whom are being subjected to imprisonment and re-education in internment camps. For someone who isn't knowledgeable in Chinese history and hasn't been following this story, Wikipedia may be their only exposure to this group's history. 


To create a training set, we scraped a set of Wikipedia pages that discussed Chinese history and fed into the \textit{MIM Engine}. In total, we scraped 19 articles, including "1989 Tiananmen Square protests,"\cite{wiki_1989_tiananmen} "Archaeology of China,"\cite{wiki_archaeology_of_china} "Chinese Communist Revolution,"\cite{wiki_chinese_communist_revolution} "Chinese economic reform,"\cite{wiki_chinese_economic_reform} "Chinese literature,"\cite{wiki_chinese_literature} "Cultural Revolution,"\cite{wiki_cultural_revolution} "Ethnic groups in Chinese history,"\cite{wiki_ethnic_groups} "Han Chinese,"\cite{wiki_han_chinese} "History of China,"\cite{wiki_history_of_china} "History of the People's Republic of China,"\cite{wiki_history_prc} "Hundred Days' Reform,"\cite{wiki_hundred_days_reform} "Korean War,"\cite{wiki_korean_war} "Kuomintang,"\cite{wiki_kuomintang} "Languages of China,"\cite{wiki_languages} "New Culture Movement,"\cite{wiki_new_culture_movement} "One country, two systems,"\cite{wiki_one_country_two_systems} "Second Sino-Japanese War,"\cite{wiki_second_sino_japanese_war} "Sino-Vietnamese War,"\cite{wiki_sino-vietnamese_war} and "Taiwanese indigenous peoples."\cite{wiki_taiwan_indig_peoples} The text from these articles was used as an input into the \textit{MIM engine}. 

We chose to scrape articles that discussed Chinese history so that any word replacements would discuss Chinese history, reducing the probability that someone trying to learn more about Uyghurs would identify the words that have been replaced. For target adversarial strategy, we used the words "Uyghur", "Uyghurs", "Uighur", and "Uighurs" (the latter two being alternate spellings for the former two). The resulting MIM object is shown in Table 3. The \textit{MIM Engine} replaced the target words from the adversarial manipulation strategt with the words "Manchus," "groups," "man," and "War." With this edits, the WikipediaBot manipulated article remains consistent, given that the edits in and of themsleves don't constitute an ill-intent content. Still, the adversary is able to eliminate any reference to the "Uyghurs" in the article, achieving their misinformation objective. 

\begin{table}[h]
\caption {Original Wikipedia Article [textual content]}
\centering
\begin{tabularx}{\linewidth}{|X|}
\hline

Despite the ongoing repression of the \textbf{Uyghurs} as portrayed by Western media, there have been very few protests from Islamic countries against the internment and re-education of the ethnicity by the Chinese Communist Party. In December 2018, the Organisation of Islamic Cooperation (OIC) initially acknowledged the disturbing reports from the region but the statement was later retracted and replaced by the comment that the OIC "commends the efforts of the People's Republic of China in providing care to its Muslim citizens; and looks forward to further cooperation between the OIC and the People's Republic of China." Even Saudi Arabia, which host significant numbers of ethnic \textbf{Uyghurs}, have refrained from any official criticism of the Chinese government, possibly due to economic and political liaisons between China and many Islamic nations.[157][158] While the Turkish Foreign Ministry spokesperson denounced China for "violating the fundamental human rights of \textbf{Uyghur} Turks and other Muslim communities in the Xinjiang \textbf{Uyghur} Autonomous Region" in February 2019,[159][160] Turkish President Erdogan later said “It is a fact that the people of all ethnicities in Xinjiang are leading a happy life amid China's development and prosperity” while visiting China.[161] Erdogan also said that some people were seeking to "abuse" the Xinjiang crisis to jeopardize Turkey and China's economic relationship.[162][163][164] \\
\hline
\end{tabularx}
\end{table}

\begin{table}[h]
\caption {WikipediaBot Manipulated Article [textual content]}
\centering
\begin{tabularx}{\linewidth}{|X|}
\hline
Despite the ongoing repression of the \textbf{Manchus} as portrayed by Western media, there have been very few protests from Islamic countries against the internment and re-education of the ethnicity by the Chinese Communist Party. In December 2018, the Organisation of Islamic Cooperation (OIC) initially acknowledged the disturbing reports from the region but the statement was later retracted and replaced by the comment that the OIC "commends the efforts of the People's Republic of China in providing care to its Muslim citizens; and looks forward to further cooperation between the OIC and the People's Republic of China." Even Saudi Arabia, which host significant numbers of ethnic \textbf{groups} have refrained from any official criticism of the Chinese government, possibly due to economic and political liaisons between China and many Islamic nations.[157][158] While the Turkish Foreign Ministry spokesperson denounced China for "violating the fundamental human rights of \textbf{man.} Turks and other Muslim communities in the Xinjiang \textbf{War} Autonomous Region" in February 2019,[159][160] Turkish President Erdogan later said “It is a fact that the people of all ethnicities in Xinjiang are leading a happy life amid China's development and prosperity” while visiting China.[161] Erdogan also said that some people were seeking to "abuse" the Xinjiang crisis to jeopardize Turkey and China's economic relationship.[162][163][164] \\
\hline
\end{tabularx}
\end{table}

\subsection{Performance} 
Due to the context in which we are performing our research, and because we are not willing to deface Wikipedia without coordination with the Wikipedia security team, we didn't proceed with last step of the execution of the MIM object through the \textit{Change Executor} module. A proper performance test, outside of sneaky vandalism detection, would also entail comparison between participants exposed to the original and the WikipediaBot manipulated articles to capture the effect in inducing misperceptions about the Uyghurs in controlled lab settings (pending IRB approval). 

Possible metrics that an adversary could use to test the subtlety of their changes would be to measure the number of edits that have happened to the page before the manipulated page has been corrected. Measuring how long the altered text remains available can help an adversary to refine their approach until they are consistently able to create misperceptions in text that are not quickly corrected. This could be measured in the future through the manipulation of pages that are less likely to affect people in a human rights context, such as by editing Wikipedia articles that affect fictional universes, e.g. \textit{Game of Thrones} character "Daenerys Targaryen," or the article for "Decepticon" from \textit{Transformers}. Researching how users in live environments interact with MIM could help to develop better understanding of how security teams could protect the integrity of the text in a target Wikipedia Article.

In addition to subtlety, an adversary could also measure effectiveness of their MIM in persuading the targeted reader to perceive text in a specific way. The simplest approach would be to use a survey with a text snippet to see how much the adversarial edits affects the interpretation of the text by an unwitting reader. Using a service like Amazon Mechanical Turk, an adversary could present a couple of quotes from their MIM engine, and ask the survey respondents questions about how they feel about the topic in question. 

\section{Detection, Evasion, and Defenses} 
\label{sec:5}
\subsection {Detection and Evasion}
The Wikipedia articles targeted by the WikipediaBot are organically written by a human. The minor edits that are part of the MIM object provide the adversary a tactical advantage because they mimic an organic editing of the articles themselves. Therefore, many of the existing tools for detecting automatically generated text will not work, and could even affirm the integrity of a  WikipediaBot manipulated text as being genuinely edited. Instead of using the bot feature in a default manner (where multiple "bot users" are logging in, going straight to the target Wikipedia page, clicking edit, then making a small change and logging out), the interactions that each bot user could be altered by adding time to different steps, by doing some browsing on related topics, or even by staying logged in and using different browsers for the other login credential sets. The adversary could also use WikipediaBot, rather than continuously running the edits.


From a human reader evasion perspective, viewing a WikipediaBot manipulated sample using the GLTR detector \cite{GLTR} at a cursory examination indicates that the text was not machine generated, as shown in Figure 4 (in this example, the word "Uyghur" was swapped with the word "Han"). 


\begin{figure}[!h]
  \centering
  \includegraphics[width=\linewidth]{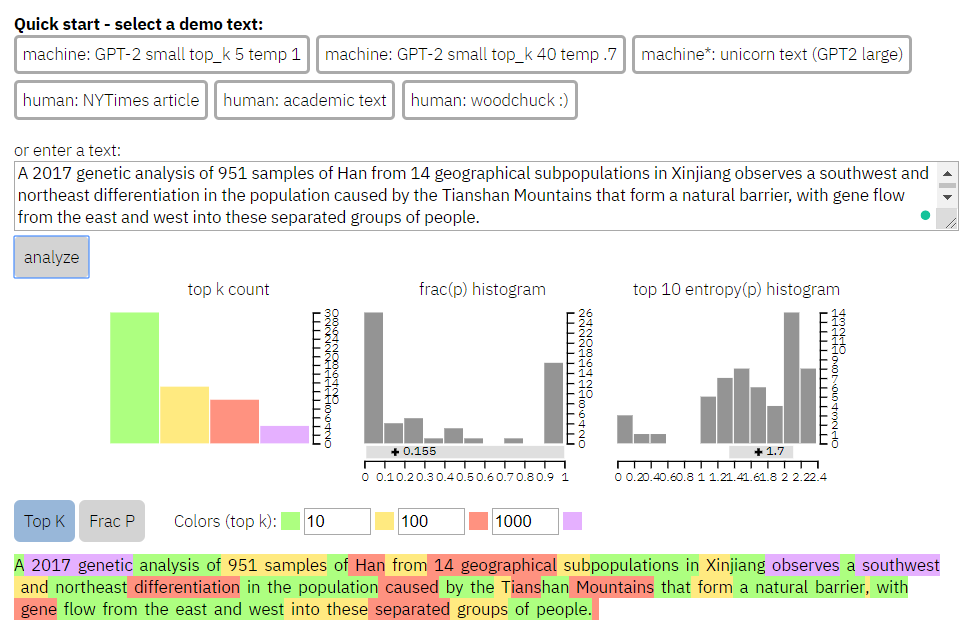}
  \caption{GLTR's Performance on a WikipediaBot manipulated article.}
\end{figure}


One feature that GLTR makes available that could help to mitigate the likelihood of successful adversarial manipulations is that a user can click on and around a possibly edited word that may look out of place to the reader as shown in Figure 5. However, the WikipediaBot edits could evade even this manual inspection. Further inspection of the words around the edited word ("Han") shows that the words around the edited word still fit in the domain of Chinese history (the Han are China's most prevalent ethnic group).

%
%

\begin{figure}[h]
  \centering
  \includegraphics[width=\linewidth]{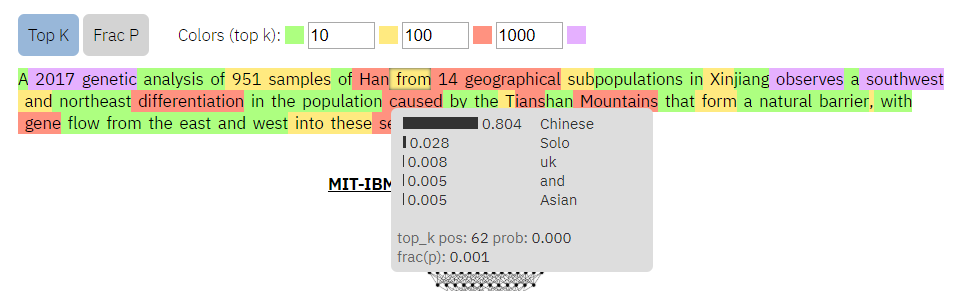}
  \caption{Details of an Adversarial Edit: Replaced Word}
\end{figure}

\subsection{Defense Against WikipediaBot}
An easy way to thwart the WikipediaBot is to add a more robust CAPTCHA system to prevent edits to individual pages. Using Google's reCAPTCHA, for example, could increase the complexity of editing pages \cite{googlecaptcha}. Even if the adversary uses Selenium to open the Wikipedia page in a browser, the reCAPTCHA will dramatically increase the complexity of the \textit{Generate Logins}, consequently reducing the usefulness of automating the entire process of adversarial manipulations. 

A possible mechanism for identifying the WikipediaBot adversarial manipulations, which is more resilient to behavior modifications by the adversary, is to use a forensics approach to try to isolate what target word strategy the adversary used, then recreating pieces of their training set from open data sources. In the example of the Uighur text replacement, once adversarial edits are identified against an earlier version of the page, the defenders could try searching for the strings in other related Wikipedia pages, e.g. searching for "samples of Han," and "of Han," etc., until they identify where is the most likely strategy of the adversary.

\section{Enhancements, Limitations, Implications}
\label{sec:4}
\subsection {Technical Enhancements}
While not necessary for this proof-of-concept, we identified two primary types of adaptations that could be used to improve the WikipediaBot mechanism should the Wikipedia security team start to increase its defenses against this type of attack, and users become cognizant of the potential for bad actors to automatically alter the meaning of Wikipedia articles. First, enhancements to the approach intended to help evade automated detection of WikipediaBot's mischief; second, improvements to the MIM object intended to improve the quality and efficiency of text alterations.

Enhancements to the approach that we identified include strategies to reduce the consistency with which the bots make their edits. For example, to avoid correlating multiple users with specific IP addresses, an adversary could use multiple computers to make updates \cite{Tran}. To reduce the likelihood of the timestamps of pauses being used to identify suspicious behavior, the pauses between actions could be timed to better mimic the time it takes a human to make an edit \cite{Rad}. To reduce the likelihood of homogeneous clickpaths being used to identify suspicious behavior, the bot accounts could be programmed to randomly follow links in the articles, and accessing multiple pages before landing on the target Wikipedia article page where it makes its edits.

Additional enhancements to the MIM object that we hope to experiment with could help to increase the subtlety and influence of our adversarial edits. Currently, our candidate strategy requires manually supplying the \textit{MIM engine} a list of target words. One possible enhancement is to use a variant of the Markov chain to determine the best candidate strategy, and make updates without supplying the list of target words. This could improve the subtlety of our \textit{MIM Engine} by making our adversarial strategy less predictable. The MIM object could further be enhanced by making sure that the candidate word has the same part of speech as the target word to help reduce the introduction of typos into the text. Additionally, the MIM object could be enhanced by equipping it with a grammar checker, that could check the grammar of an edit to make sure that it is grammatically correct. The grammar checker could also be used to check the effectiveness of an adversarial strategy by comparing the average number of grammatical errors introduced per word substitution, and this could be used to help the adversary more quickly develop their candidate strategy. The grammar checker could be used to make non-adversarial edits to pages to reduce the likelihood of the account being identified as a vandal.

\subsection {Limitations}
The automated adversarial manipulations using the WikipediaBot have several limits. Regardless of the predictive potential of the MIM engine, if the adversary uses an ill-defined strategy, for example target word replacements that make no sense, the WikipediaBot is unable to compensate for the common sense logic and the actual readability of the manipulations. Another limitation is that the WikipediaBot might behave differently when targeting different articles on Wikipedia. The WikipediaBot edits might not in and of themselves be detectable, but nontheless be detected due to previous edits for which the adversary was not aware (that might have placed a particular article as a high target of sneaky Wikipedia vandalism). Additionally, a selection of other adversarial machine learning approaches for generating target edits then the Markov Chain MIM approach might result into a different MIM output than the one produced by the WikipediaBot. 

\subsection {Ethical Implications}
When automated mechanisms like WikipediaBot take on adversarial work with social dimensions, the introduction to the wider public inherits the social requirements for well-being of the Wikipedia ecosystem and responsible disclosure \cite{Frankish}. By transparently sharing the paradigm and the proof-of-concept of the WikipediaBot mechanism, we believe that the adversarial advantage can be exposed, and with that, ultimately make the Wikipedia moderators aware of this distributed sneaky vandalism approach. To this objective, we contacted the Wikipedia security team with the details and the inner workings of WikipediaBot prior to writing this publication as part of the responsible disclosure requirement. The exposure of the WikipediaBot system architecture allows for consideration of  other types of detection, prevention, and defenses then the one proposed in this paper. We only tested the WikipediaBot on a local, isolated testbed, and never used it to make any adversarial manipulation on the live Wikipedia platform. A full scale analysis of the WikipediaBot would require coordinated work with the Wikipedia security team.   

\section {Conclusion}
\label{sec:6}
In this work, we introduced WikipediaBot, a mechanism that allows an adversary to create and control a bot infrastructure for the purpose of automated manipulation of Wikipedia articles. The goal of these adversarial edits, deemed as \textit{sneaky vandalism}, is to induce misperception withing the general public to the goal of the adversary by compromising the way the Wikipedia article is interpreted by normal, unwitting readers. We believe that considering automated mechanisms such as the WikipediaBot from a research perspective is an important step towards understating the means, motives, and opportunities of adversaries interested in spreading disinformation, rumors, trolling content, and fake content through alternative means. We hope our work informs the security community about the potential of automatic adversarial manipulation of textual content online. 


\bibliographystyle{plain}
\bibliography{WikipediaBot.bib}

\end{document}